\documentclass[a4paper]{article}
\usepackage{amsmath}
\usepackage{epsfig}
\usepackage{natbib}
\usepackage{amssymb}

\begin{document}

\title{Competition between synaptic depression and facilitation in attractor
neural networks}
\author{J. J. Torres$^{\dag }$, J.M. Cortes$^{\dag \ddag}$\footnote{Present address: Institute for Adaptive and Neural Computation, School of
Informatics, University of Edinburgh, 5 Forrest Hill, EH1 2QL, UK.}%
\thinspace , J. Marro$^{\dag }$ and H.J. Kappen$^{\ddag }$ \\
$^{\dag}$Institute \textit{Carlos I} for Theoretical and Computational
Physics, and \\
Departamento de Electromagnetismo y F\'{\i}sica de la Materia,\\
University of Granada, E-18071 Granada, Spain.\\
$^{\ddag}$Department of Biophysics and SNN, Radboud University of Nijmegen, \\
6525 EZ Nijmegen, The Netherlands}
\maketitle

\begin{abstract}
We study the effect of competition between short-term synaptic depression
and facilitation on the dynamical properties of attractor neural
networks, using Monte Carlo simulation and a mean field analysis.
Depending on the balance between depression, facilitation and the noise, the
network displays different behaviours, including
associative \textit{memory} and 
switching of the activity between different
attractors. We conclude that synaptic facilitation enhances the
attractor instability in a way that (\textit{i}) intensifies the system
adaptability to external stimuli, which is in agreement with experiments,
and (\textit{ii}) favours the retrieval of information with less error
during short--time intervals.

\end{abstract}

\section{Introduction and model}
Recurrent neural networks are a prominent model for information
processing and memory in the brain. 
\citep{hopfield,amitB}. Traditionally, these models assume 
synapses that may change on the time scale of learning, but that can
be assumed constant during memory retrieval.
However, synapses are reported to exhibit rapid time variations, and 
it is likely that this finding has important implications for our 
understanding of the way information is processed in the brain
\citep{abbNATURE}. For instance, Hopfield--like networks in which
synapses undergo rather generic fluctuations have been shown to
significantly improve the associative process, e.g., \citep{marroPRL}.
In addition, motivated by specific neurobiological observations and their
theoretical interpretation \citep{tsodyksNC}, activity--dependent synaptic
changes which induce \textit{depression} of the response have been
considered \citep{torresNC,bibitchkov}. It was shown
that synaptic depression induces, in addition to memories as stable
attractors, special sensitivity of the network to
changing stimuli as well as rapid switching of the activity among the
stored patterns %
\citep{torresNC,cortesNEUCOM,marroPRE,cortesBIOPHYSCHEM,cortesNC}.
This behaviour 
has been observed experimentally to occur during the processing of sensory
information \citep{laurentANNUREVNEUROSCI,mazor,hibrido}.

In this paper, we present and study
networks that are inspired in the observation of certain, more complex
synaptic changes. That is, we assume that repeated presynaptic activation
induces at short times not only depression but also \textit{facilitation} of
the postsynaptic potential \citep{thomsonTRENDSNEUROSCI,facil,facil2}. The
question, which has not been quite addressed yet, is how a competition
between depression and facilitation will affect the network performance. We
here conclude that, as for the case of only depression %
\citep{torresNC,cortesNC}, the system may exhibit up to three different 
\textit{phases} or regimes, namely, one with standard associative memory, a
disordered phase in which the network lacks this property, and an
oscillatory phase in which activity switches between different memories.
Depending on the balance between facilitation and depression, novel intriguing behavior
results in the oscillatory regime. In particular, as the degree of
facilitation increases, both the sensitivity to external stimuli is enhanced
and the frequency of the oscillations increases. It then follows that
facilitation allows for recovering of information with less error, at least
during a short interval of time
and can therefore play an important role in short--term memory processes.
We are concerned in this paper with a network of binary
neurons. Previous studies have shown that the
behaviour of such a simple network dynamics agree qualitatively with
the behaviour that is observed in more realistic networks, such as
integrate and fire neuron models of pyramidal cells~\citep{torresNC}.

Let us consider $N$ binary neurons, $s_{i}=\{1,0\},$ $i=1,...,N,$ endowed of
a probabilistic dynamics, namely,%
\begin{equation}
\mathrm{Prob}\left\{ s_{i}\left( t+1\right) =1\right\} =\frac{1}{2}\left\{
1+\tanh \left[ 2\beta h_{i}\left( t\right) \right] \right\} ,  \label{s}
\end{equation}%
which is controlled by a \textit{temperature} parameter, $T\equiv 1/\beta ;$
see, for instance, \citep{marroB} for details. The function $h_{i}\left(
t\right) $ denotes a time--dependent \textit{local field}, i.e., the total
presynaptic current arriving to the postsynaptic neuron $i.$ This will be
determined in the model following the phenomenological description of
nonlinear synapses reported in \citep{markramPNAS,tsodyksNC}, which was
shown to capture well the experimentally observed properties of neocortical
connections. Accordingly, we assume that 
\begin{equation}
h_{i}\left( t\right) =\sum_{j=1}^{N}\omega _{ij}\mathcal{D}_{j}\left(
t\right) \mathcal{F}_{j}\left( t\right) s_{j}\left( t\right) -\theta _{i},
\label{h}
\end{equation}%
where $\theta _{i}$ is a constant threshold associated to the firing of
neuron $i,$ and $\mathcal{D}_{j}(t)$ and $\mathcal{F}_{j}(t)$ are functions
---to be determined--- which describe the effect on the neuron activity of
short--term synaptic depression and facilitation, respectively. We further
assume that the weight $\omega _{ij}$ of the connection between the
(presynaptic) neuron $j$ and the (postsynaptic) neuron $i$ are static and 
\textit{store} a set of patterns of the network activity, namely, the
familiar \textit{covariance rule}: 
\begin{equation}
\omega _{ij}=\frac{1}{Nf\left( 1-f\right) }\sum_{\nu =1}^{P}\left( \xi
_{i}^{\nu }-f\right) \left( \xi _{j}^{\nu }-f\right) .
\end{equation}%
Here, $\mathrm{\xi }^{\nu }=\left\{ \xi _{i}^{\nu }\right\} ,$ with $\nu
=1\ldots ,P,$ are different binary--patterns of average activity $\langle
\xi _{i}^{\nu }\rangle \equiv f$. The standard Hopfield model is recovered
for $\mathcal{F}_{j}=\mathcal{D}_{j}=1$, $\forall j=1,\ldots ,N.$

We next implement a dynamics for $\mathcal{F}_{j}$ and $\mathcal{D}_{j}$
after the prescription in \citep{markramPNAS,tsodyksNC}. A description of
varying synapses requires, at least, three local variables, say $x_{j}\left(
t\right) $, $y_{j}\left( t\right) $ and $z_{j}\left( t\right) ,$ to be
associated to the fractions of neurotransmitters in recovered, active, and
inactive states, respectively. A simpler picture consists in dealing with
only the $x_{j}\left( t\right) $ variable. This simplification, which seems
to describe accurately both interpyramidal and pyramidal interneuron
synapses, corresponds to the fact that the time in which the postsynaptic
current decays is much shorter than the recovery time for synaptic
depression, say $\tau _{\mathrm{rec}}$ \citep{markramNATURE} (Time intervals are in milliseconds hereafter). Within this
approach, one may write that 
\begin{equation}
x_{j}\left( t+1\right) =x_{j}\left( t\right) +\frac{1-x_{j}\left( t\right) }{%
\tau _{\mathrm{rec}}}-\mathcal{D}_{j}\left( t\right) \mathcal{F}_{j}\left(
t\right) s_{j}\left( t\right) ,  \label{x}
\end{equation}%
where%
\begin{equation}
\mathcal{D}_{j}\left( t\right) =x_{j}\left( t\right)  \label{D}
\end{equation}%
and%
\begin{equation}
\mathcal{F}_{j}\left( t\right) =U+\left( 1-U\right) \,u_{j}\left( t\right) .
\label{F}
\end{equation}%
The interpretation of this ansatz is as follows. Concerning any presynaptic
neuron $j,$ the product $\mathcal{D}_{j}\mathcal{F}_{j}$ stands for the
total fraction of neurotransmitters in the recovered state which are
activated either by incoming spikes, $U_{j}x_{j},$ or by facilitation
mechanisms, $\left( 1-U_{j}\right) x_{j}u_{j};$ for simplicity, we are
assuming that $U_{j}=U\in \left[ 0,1\right] $ $\forall j.$ The additional
variable $u_{j}\left( t\right) $ is assumed to satisfy, as in the quantal
model of transmitter release in \citep{markramPNAS}, that%
\begin{equation}
u_{j}\left( t+1\right) =u_{j}\left( t\right) -\frac{u_{j}\left( t\right) }{%
\tau _{\mathrm{fac}}}+U\left[ 1-u_{j}\left( t\right) \right] s_{j}\left(
t\right) ,  \label{u}
\end{equation}%
which describes an increase with each presynaptic spike and a decay to the
resting value with relaxation time $\tau _{\mathrm{fac}}$ (that is given in milliseconds) 
Consequently,
facilitation washes out ($u_j \rightarrow 0, F_j \rightarrow U$) as 
$\tau _{\mathrm{fac}}\rightarrow 0,$ and each
presynaptic spike uses a fraction $U$ of the available resources 
$x_{j}\left( t\right) .$ The effect of facilitation increases with decreasing 
$U,$ because this will leave more neurotransmitters available to be
activated by facilitation. Therefore, facilitation is not controlled only by 
$\tau _{\mathrm{fac}}$ but also by $U.$

The Hopfield case with static synapses is recovered after using $x_{j}=1$ in
eq.(\ref{D}) and $u_{j}=0$ in eq.(\ref{F}) or, equivalently, $\tau _{\mathrm{%
rec}}=\tau _{\mathrm{fac}}=0$ in eqs. (\ref{x}) and (\ref{u}). In fact, the
latter imply fields $h_{i}\left( t\right) =\sum_{j}\omega _{ij}Us_{j}\left(
t\right) -\theta _{i},$ so that one may simply rescale both $\beta $ and the
threshold.

The above interesting phenomenological description of dynamic changes has
already been implemented in attractor neural networks \citep{torresNC} for
pure depressing synapses between excitatory pyramidal neurons %
\citep{tsodyksPNAS}. We are here interested in the consequences of a
competition between depression and facilitation. Therefore, we shall use $%
T,U,\tau _{\mathrm{rec}}$ and $\tau _{\mathrm{fac}}$ in the following as
relevant control parameters.

\section{Mean--field solution}

Let us consider the mean activities associated, respectively, with active
and inert neurons in a given pattern $\nu ,$ namely,%
\begin{equation}
m_{+}^{\nu }(t)\equiv \frac{1}{Nf}\sum_{j\in \text{Act}(\nu )}s_{j}(t),\quad
m_{-}^{\nu }(t)\equiv \frac{1}{N(1-f)}\sum_{j\not\in \text{Act}(\nu
)}s_{j}(t).
\end{equation}%
It follows for the overlap of the network activity with pattern $\nu $ that%
\begin{equation}
m^{\nu }(t)\equiv \frac{1}{Nf\left( 1-f\right) }\sum_{i}\left( \xi _{i}^{\nu
}-f\right) s_{i}(t)=m_{+}^{\nu }(t)-m_{-}^{\nu }(t),  \label{m}
\end{equation}%
$\forall \nu .$ One may also define the averages of $x_{i}$ and $u_{i}$ over
the sites that are active and inert, respectively, in a given pattern $\nu ,$
namely,%
\begin{equation}
\begin{array}{c}
x_{+}^{\nu }(t)\equiv \dfrac{1}{Nf}\sum_{j\in \text{Act}(\nu
)}x_{j}(t),\quad x_{-}^{\nu }(t)\equiv \dfrac{1}{N(1-f)}\sum_{j\not\in \text{%
Act}(\nu )}x_{j}(t) \\ 
u_{+}^{\nu }(t)\equiv \dfrac{1}{Nf}\sum_{j\in \text{Act}(\nu
)}u_{j}(t),\quad u_{-}^{\nu }(t)\equiv \dfrac{1}{N(1-f)}\sum_{j\not\in \text{%
Act}(\nu )}u_{j}(t),%
\end{array}%
\end{equation}%
$\forall \nu ,$ which describe depression (the $x$s) and facilitation (the $%
u $s), each concerning a subset of neurons, e.g., $N/2$ neurons for $f=1/2.$
The local fields then ensue as 
\begin{equation}
h_{i}(t)=\sum_{\nu =1}^{P}\left( \xi _{i}^{\nu }-f\right) M^{\nu }\left(
t\right) ,  \label{haprox}
\end{equation}%
$M^{\nu }(t)\equiv \left[ U+\left( 1-U\right) \,u_{+}^{\nu }(t)\right]
\,x_{+}^{\nu }(t)\,m_{+}^{\nu }(t)-\left[ U-\left( 1-U\right) \,u_{-}^{\nu
}(t)\right] \,x_{-}^{\nu }(t)\,m_{-}^{\nu }(t).$

One may solve the model (\ref{s})--(\ref{u}) in the thermodynamic limit $%
N\rightarrow \infty $ under the standard mean-field assumption that $%
s_{i}\approx \langle s_{i}\rangle .$ Within this approximation, we may also
substitute $x_{i}$ ($u_{i}$) 
by the mean--field values $x_{\pm }^{\nu }$ ($u_{\pm }^{\nu }$). (Notice
that one expects, and it will be confirmed below by comparisons with direct
simulation results, that the mean--field approximation is accurate away from
any possible critical point.) Assuming further that patterns are random with
mean activity $f=1/2,$ one obtains the set of dynamic equations:%
\begin{equation*}
\qquad \quad x_{\pm }^{\nu }(t+1)=x_{\pm }^{\nu }(t)+\frac{1-x_{\pm }^{\nu
}(t)}{\tau _{\mathrm{rec}}}-\left[ U+\left( 1-U\right) \,u_{\pm }^{\nu }(t)\,%
\right] \,x_{\pm }^{\nu }(t)\,m_{\pm }^{\nu }(t),
\end{equation*}%
\begin{equation*}
u_{\pm }^{\nu }(t+1)=u_{\pm }^{\nu }(t)-\frac{u_{\pm }^{\nu }(t)}{\tau _{%
\mathrm{fac}}}+U\,[1-u_{\pm }^{\nu }(t)]\,m_{\pm }^{\nu }(t),\qquad \qquad
\end{equation*}%
\begin{equation*}
\qquad m_{\pm }^{\nu }(t+1)=\frac{1}{N}{\sum_{i}}\left\{ 1\pm \tanh \left(
\beta \left[ M^{\nu }\left( t\right) \pm \sum_{\mu \neq \nu }\epsilon
_{i}^{\mu }M^{\mu }\left( t\right) \right] \right) \right\} ,
\end{equation*}%
\begin{equation}
m^{\nu }(t+1)=\frac{1}{N}\sum_{i}\epsilon _{i}^{\nu }\tanh \left[ \beta
\sum_{\mu }\epsilon _{i}^{\mu }M^{\mu }\left( t\right) \right] ,\qquad
\qquad \qquad  \label{dyn}
\end{equation}%
where $\epsilon _{i}^{\mu }\equiv 2\xi _{i}^{\mu }-1.$ This is a $6P$%
--dimensional coupled map whose analytical treatment is difficult for large $%
P,$ but it may be integrated numerically, at least for not too large $P.$
One may also find the fixed--point equations for the coupled dynamics of
neurons and synapses; these are%
\begin{eqnarray}
x_{\pm }^{\nu } &=&\left\{ 1+\,\left[ U\,+\,\left( 1-U\right) \,\,u_{\pm
}^{\nu }\right] \,\tau _{\mathrm{rec}}\,m_{\pm }^{\nu }\right\} ^{-1}, 
\notag \\
u_{\pm }^{\nu } &=&U\,\,\tau _{\mathrm{fac}}\,\,m_{\pm }^{\nu }\left(
1\,\,+\,\,U\,\,\tau _{\mathrm{fac}}\,\,m_{\pm }^{\nu }\right) ^{-1},  \notag
\\
2m_{\pm }^{\nu } &=&1\pm \frac{2}{N}{\sum_{i}}\tanh \left[ \beta \left(
M^{\nu }\pm \sum_{\mu \neq \nu }\epsilon _{i}^{\mu }M^{\mu }\right) \right] ,
\notag \\
m^{\nu } &=&\frac{1}{N}\sum_{i}\epsilon _{i}^{\nu }\tanh \left( \beta
\sum_{\mu }\epsilon _{i}^{\mu }M^{\mu }\right) .  \label{steady}
\end{eqnarray}%
The numerical solution of these transcendental equations describes the
resulting order as a function of the relevant parameters. Determining the
stability of these solutions for $\alpha =P/N\neq 0$ is a more difficult
task, because it requires to linearize (\ref{dyn}) and the dimensionality
diverges in the thermodynamical limit (see however~\citep{torresCAPACITY}). 
In the next section we
therefore deal with the case $\alpha \rightarrow 0.$

\section{Main results}

Consider a finite number of stored patterns $P,$ i.e., $\alpha
=P/N\rightarrow 0$ in the thermodynamic limit. In practice, it is sufficient
to deal with $P=1$ to illustrate the main results (therefore, we shall
suppress the index $\nu $ hereafter). 

Let us define the vectors of order
parameters $\vec{y}\equiv
(m_{+},m_{-},x_{+},x_{-},u_{+},u_{-})$, its stationary value
$\vec{y}_{st}$ that is given by the solution of Eq.~\ref{steady}, and
$\vec{F}$ whose components are the functions on the right hand side of (\ref{dyn}). 
The stability of (\ref{dyn}) around the steady state (\ref{steady}) follows
from the first derivative matrix $D\equiv \left( \partial \vec{F}\diagup
\partial \vec{y}\right) _{\vec{y}_{\mathrm{st}}}.$
This is 
\begin{equation}
D=\left( 
\begin{array}{cccccc}
\bar{\beta}A_{+} & -\bar{\beta}A_{-} & \bar{\beta}B_{+} & -\bar{\beta}B_{-}
& \bar{\beta}C_{+} & -\bar{\beta}C_{-} \\ 
-\bar{\beta}A_{+} & \bar{\beta}A_{-} & -\bar{\beta}B_{+} & \bar{\beta}B_{-}
& -\bar{\beta}C_{+} & \bar{\beta}C_{-} \\ 
-A_{+} & 0 & \tau -B_{+} & 0 & -C_{+} & 0 \\ 
0 & -A_{-} & 0 & \tau -B_{-} & 0 & -C_{-} \\ 
D_{+} & 0 & 0 & 0 & \tau -E_{+} & 0 \\ 
0 & D_{-} & 0 & 0 & 0 & \tau -E_{-}%
\end{array}%
\right)  \label{mD}
\end{equation}%
where $\bar{\beta}\equiv 2\beta m_{+}m_{-},$ $A_{\pm }\equiv \left[ U+\left(
1-U\right) \,u_{\pm }\right] \,x_{\pm },$ $B_{\pm }\equiv \left[ U+\left(
1-U\right) \,u_{\pm }\right] \,m_{\pm },$ $C_{\pm }\equiv \left( 1-U\right)
\,x_{\pm }m_{\pm },$ $D_{\pm }\equiv U(1-u_{\pm }),$ $\tau \equiv 1-\tau _{%
\mathrm{rec}}^{-1},$ and $E_{\pm }\equiv Um_{\pm }.$ After noticing that $%
m_{+}+m_{-}=1,$ one may numerically diagonalize $D$ and obtain the
eigenvalues $\lambda _{n}$ for a given set of control parameters $T,U,\tau _{%
\mathrm{rec}},\tau _{\mathrm{fac}}.$ For $\left\vert \lambda _{n}\right\vert
<1$ $(\left\vert \lambda _{n}\right\vert >1),$ the system is stable
(unstable) close to the fixed point $y_{n}$. The maximum of $\left\vert
\lambda _{n}\right\vert $ determines the local stability: for $\left\vert
\lambda _{n}\right\vert _{\mathrm{max}}<1,$ the system (\ref{dyn}) is
locally stable, while for $\left\vert \lambda _{n}\right\vert _{\mathrm{max}%
}>1$ there is at least one direction of instability, and the system
consequently becomes locally unstable. Therefore, varying the control
parameters one crosses the line $|\lambda |_{\mathrm{max}}=1$ that signals
the bifurcation points.
\begin{figure}[ht!]
\centerline{\psfig{file=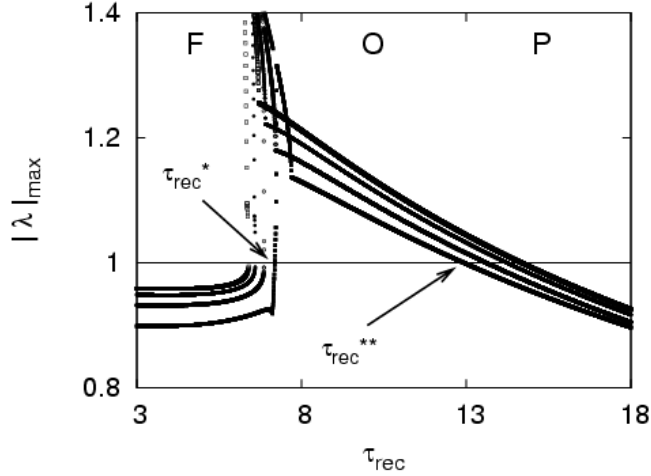,width=9.1cm}}
\caption{The
three relevant regions, denoted F, O and P, respectively, that are depicted
by the absolute value of the maximum eigenvalue $\left\vert \protect\lambda %
_{n}\right\vert _{\mathrm{max}}$ of the stability matrix $D$ in (\protect\ref%
{mD}) when plotted as a function of the recovering time $\protect\tau _{%
\mathrm{rec}}$ for different values of the facilitation time $\protect\tau _{%
\mathrm{fac}}.$ Here, $\protect\tau _{\mathrm{fac}}=10,$ 15, 20 and 25 for
different curves from bottom to top, respectively, in the F and P regions.
The stationary solutions lack of any local stability for $\protect\tau _{%
\mathrm{rec}}^{\ast }<\protect\tau _{\mathrm{rec}}<\protect\tau _{\mathrm{rec%
}}^{\ast \ast }$ (O), and the network activity then undergoes oscillations.
The arrows signal $\protect\tau _{\mathrm{rec}}^{\ast }$ and $\protect\tau _{%
\mathrm{rec}}^{\ast \ast }$ for $\protect\tau _{\mathrm{fac}}=10.$ This
graph is for $U=0.1$ and $T=0.1.$}
\label{fig1}
\end{figure}

\begin{figure}[tbh]
\centerline{\psfig{file=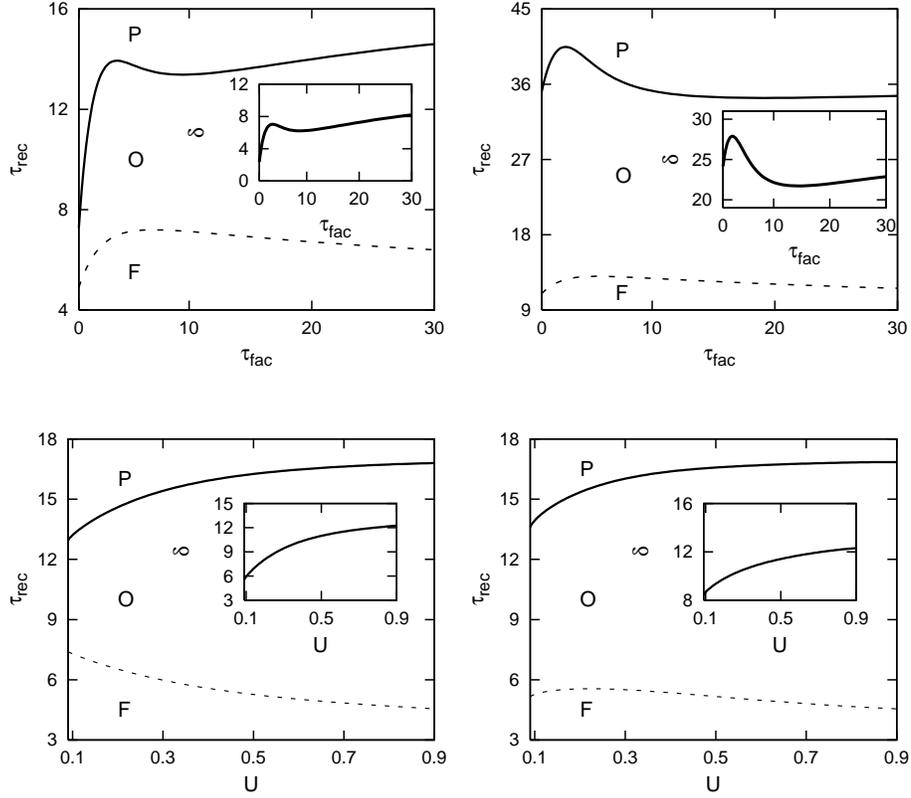,width=12cm} }
\caption{This illustrates how the different regimes of the network activity
depend on the balance between depression and facilitation. \protect%
\underline{Top graphs}: Phase diagram ($\protect\tau _{\mathrm{rec}},\protect%
\tau _{\mathrm{fac}}$) for $\protect\alpha =0$ and $U=0.1$ at temperature $%
T=0.1$ (left) and $0.05$ (right). The dashed (solid) line is for $\protect%
\tau _{\mathrm{rec}}^{\ast }$ ($\protect\tau _{\mathrm{rec}}^{\ast \ast }$)
signaling the first--order (second--order) phase transitions between the O
and F(P) phases. The insets show the resulting width of the oscillatory
region, $\protect\delta \equiv \protect\tau _{\text{rec}}^{\ast \ast }-%
\protect\tau _{\text{rec}}^{\ast },$ as a function of $\protect\tau _{%
\mathrm{fac}}.$ \protect\underline{Bottom graphs}: Phase diagram ($\protect%
\tau _{\mathrm{rec}},U$) for $\protect\alpha =0$ and $T=0.1,$ and $\protect%
\gamma \equiv \protect\tau _{\mathrm{fac}}/\protect\tau _{\mathrm{rec}}=1$
(left) and $0.25$ (right).}
\label{fig2}
\end{figure}
The resulting situation is summarized in figure \ref{fig1} for
specific values of $U,$ $T$ and $\tau _{%
\mathrm{fac}},$. Eqs.~(\ref{steady}) have three solutions, two of which
are memory states corresponding to the pattern and anti-pattern and the other a 
so-called paramagnetic state that has
no overlap with the memory pattern. The stability of the two solutions 
depends on $\tau_{\mathrm{rec}}$. 
The region 
$\tau_{\mathrm{rec}} > \tau_{rec}^{\ast\ast}$ corresponds to the non-retrieval phase,
where the paramagnetic solution is stable and the memory solutions are
unstable. In this phase, 
the average network behaviour has no significant overlap with the stored memory
pattern. The region 
$\tau_{\mathrm{rec}} < \tau_{\mathrm{rec}}^*$ corresponds to the memory phase, 
where the paramagnetic solution is unstable and the memory solutions
are stable. The network retrieves one of the stored memory patterns. 
For $\tau_{\mathrm{rec}}^{\ast }<\tau _{\mathrm{rec}}<\tau _{\mathrm{%
rec}}^{\ast \ast }$ 
(denoted ``O'' in the figure) none of the solutions is stable.
The activity of the
network in this regime keeps moving from one to the other fixed--points
neighborhood (the pattern and anti-pattern in this simple example).
This rapid switching behaviour is typical for dynamical synapses and
does not occur for static synapses. 
A similar oscillatory behavior was reported in \citep{torresNC,cortesNEUCOM} for the
case of only synaptic depression. A main novelty is that the inclusion of
facilitation importantly modifies the \textit{phase diagram}, as discussed
below (figure \ref{fig2}). On the other hand, the phases for $\tau _{\mathrm{%
rec}}<\tau _{\mathrm{rec}}^{\ast }$ (F) and $\tau _{\mathrm{rec}}>\tau _{%
\mathrm{rec}}^{\ast \ast }$ (P) correspond, respectively, to a
locally--stable regime with associative memory ($m\neq 0$) and to a
disordered regime without memory (i.e., $m\equiv m^{1}=0$).

The values $\tau _{\mathrm{rec}}^{\ast }$ and $\tau _{\mathrm{rec}}^{\ast
\ast }$ which, as a function of $\tau _{\mathrm{fac}},$ $U$ and $T,$
determine the limits of the oscillatory phase correspond to the onset of
condition $|\lambda _{n}|_{\mathrm{max}}>1.$ This condition defines lines in
the parameter space ($\tau _{\text{rec}},\tau _{\text{fac}}$) that are
illustrated in figure \ref{fig2}. This reveals that $\tau _{\text{rec}%
}^{\ast }$ (separation between the F and O regions) in general decreases
with increasing facilitation, which implies 
a larger oscillatory region and consequently a reduction of the memory phase. On the other hand, $%
\tau _{\text{rec}}^{\ast \ast }$ (separation between O and P regions) in
general increases with facilitation, thus broadening further the width of
the oscillatory phase $\delta \equiv \tau _{\text{rec}}^{\ast \ast }-\tau _{%
\text{rec}}^{\ast }.$ The behavior of this quantity under different
conditions is illustrated in the insets of figure \ref{fig2}. 

Another interesting consequence of facilitation are the changes in the 
phase diagram as one varies the facilitation parameter $U$ which measures 
the fraction of neurotransmitter
that are not activated by the facilitating mechanism\textbf{.} In order to
discuss this, we define the ratio between the time scales, $\gamma \equiv
\tau _{\mathrm{fac}}/\tau _{\mathrm{rec}},$ and monitor the phase diagram ($%
\tau _{\mathrm{rec}},U$) for varying $\gamma .$ The result is also in figure %
\ref{fig2} ---see the bottom graphs for $\gamma =1$ (left) and $0.25$
(right) which correspond, respectively, to a situation in which depression
and facilitation occur in the same time scale and to a situation in which
facilitation is four times faster. The two cases exhibit a similar behavior
for large $U,$ but they are qualitatively different for small $U.$ In the
case of faster facilitation, there is a range of $U$ values for which $\tau
_{\text{rec}}^{\ast }$ increases, in such a way that one passes from the
oscillatory to the memory phase by slightly increasing $U.$ This means that
facilitation tries to drive the network activity to one of the attractors ($%
\tau _{\mathrm{fac}}<\tau _{\mathrm{rec}}$) and, for weak depression ($U$
small), the activity will remain there. Decreasing $U$ further has then the
effect of increasing effectively the system temperature, which destabilizes
the attractor. This only requires small $U$ because the dynamics (\ref{u})
rapidly decreases the second term in $\mathcal{F}_{j}$ to zero.

\begin{figure}[tbh]
\centerline{
\psfig{file=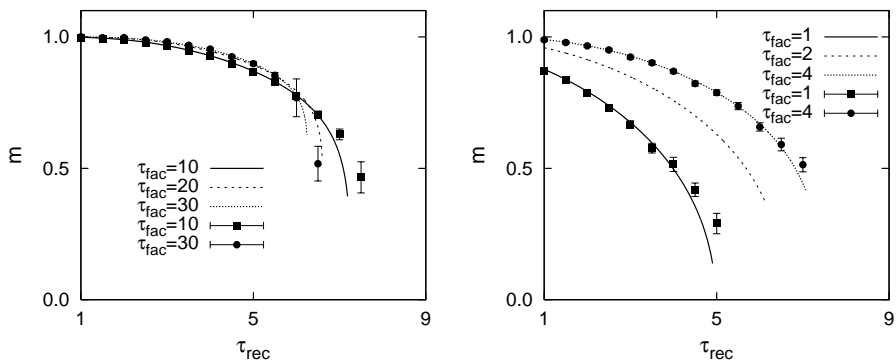,width=12cm}
}
\caption{For $U=T=0.1$ as in figure \protect\ref{fig1}, these graphs
illustrate results from Monte Carlo simulations (symbols) and mean--field
solutions (curves) for the case of associative memory under competition of
depression and facilitation. This shows $m\equiv m^{1}$ as a function of $%
\protect\tau _{\mathrm{rec}}$ (horizontal axis) and $\protect\tau _{\mathrm{%
fact}}$ (different curves as indicated) corresponding to regimes in which
the limiting value $\protect\tau _{\mathrm{rec}}^{\ast }$ decreases (left
graph) or increases (right graph) with increasing $\protect\tau _{\mathrm{fac%
}},$ the two situations that are discussed in the main text.}
\label{fig3}
\end{figure}

Figure \ref{fig3} shows the variation with both $\tau _{\mathrm{rec}}$ and $%
\tau _{\mathrm{fac}}$ of the stationary locally--stable solution with
associative memory, $m\neq 0$, computed this time both in the mean field 
approximation and using Monte Carlo simulation. This Monte Carlo simulation 
consists of iterating eqs. (\ref{s}), (\ref{x}) and (\ref{u}) using parallel dynamics.
This shows a perfect agreement between our
mean--field approach above and Monte Carlo simulations as long as one is far
from the transition, a fact which is confirmed below (in figure \ref{figure5}%
). This is because, near $\tau _{\mathrm{rec}}^{\ast },$ the simulations
describe hops between positive and negative $m$ which do not compare well
with the mean--field absolute value $\left\vert m\right\vert .$

\begin{figure}[tbh]
\centerline{
\psfig{file=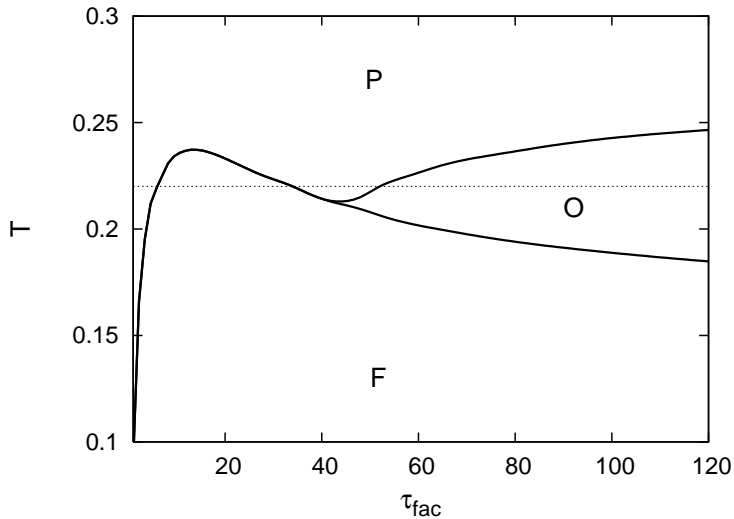,width=10cm}  
}
\caption{Phase diagram $(T,\protect\tau _{\mathrm{fac}})$ for $U=0.1$ and $%
\protect\tau _{\mathrm{rec}}=3$ms. This illustrates the potential high
adaptability of the network to different tasks, e.g., around $T=0.22,$ by
simply varying its degree of facilitation.}
\label{figure4}
\end{figure}

The most interesting behavior is perhaps the one revealed by the phase
diagram $(T,\tau _{\mathrm{fac}})$ in figure \ref{figure4}. Here we depict a
case with $U=0.1,$ in order to clearly visualize the effect of facilitation
---facilitation has practically no effect for any $U>0.5,$ as shown above---
and $\tau _{\mathrm{rec}}=3$ms in order to compare with the situation of
only depression in \cite{torresNC}. A main result here is that, for
appropriate values of the working temperature $T$, one may force the system
to undergo different types of transitions by simply varying $\tau _{\mathrm{%
fac}}.$ First note, that the line $\tau_{\mathrm{fac}}=0$ corresponds roughly to the
case of static synapses, since $\tau_{\mathrm{rec}}$ is very small. In this limit the transition
between retrieval (F) and non-retrieval (P) phases is at
$T=U=0.1$ At low enough $T,$ there is transition between the non--retrieval (P)
and retrieval phases (F) as facilitation is increased. This reveals a
positive effect of facilitation on memory at low temperature, and suggests
improvement of the network storage capacity which is usually measured at $%
T=0,$ a prediction that we have confirmed in preliminary simulations. At
intermediate temperatures, e.g., $T\approx 0.22$ for $U=0.1,$ the systems
shows no memory in the absence of facilitation, but increasing $\tau _{%
\mathrm{fac}}$ one may describe consecutive transitions to a retrieval phase
(F), to a disordered phase (P), and then to an oscillatory phase (O). The
latter is associated to a new instability induced by a strong depression
effect due to the further increase of facilitation. At higher $T,$
facilitation may drive the system directly from complete disorder to an
oscillatory regime.

\begin{figure}[tbh]
\centerline{
\psfig{file=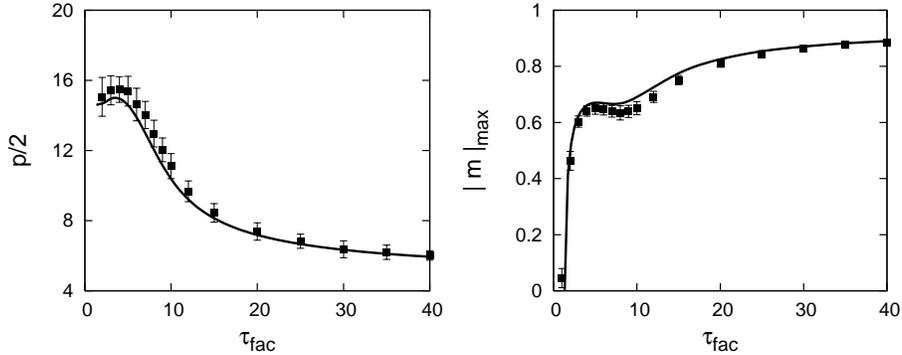,width=12cm}
}
\caption{\protect\underline{{Left graph}}{: Half period of oscillations as a
function of $\protect\tau _{\text{fac}}$, for $\protect\tau _{\text{rec}%
}=10, $ $U=T=0.1$ and $P=1,$ as obtained from the mean--field solution
(solid curve) and from simulations (symbols). \protect\underline{{Right graph%
}}: For the same conditions than in the left graph, this shows the maximum
of the absolute value of $m$ during oscillations. The simulation results in
both graphs correspond to an average over $10^{3}$ peaks of the stationary
series for $m$. The fact the statistical errors are small confirms a
periodic behavior.}}
\label{figure5}
\end{figure}

In addition to its influence on the onset and width of the oscillatory
region, $\tau _{\text{fac}}$ determines the frequency of the oscillations of 
$m.$ In order to study this effect, we computed the average time between
consecutive minimum and maximum of these oscillations, i.e., a half period.
The result is illustrated in the left graph of figure \ref{figure5}. {This
shows that the frequency of the oscillations increases with the facilitation
time. This means that the access of the network activity to the attractors
is faster with increasing facilitation, though the system then remains a
shorter time near each attractor due to an stronger depression. }On the
other hand, we also computed the maximum of $m$ during oscillations, namely, 
$|m|_{\mathrm{max}}.$ This, which is depicted in the right graph of figure %
\ref{figure5}, also increases with $\tau _{\text{fac}}.$ The overall
conclusion is that not only the access to the stored information is faster
under facilitation but that {increasing facilitation will also help to
retrieve information with less error.}

\begin{figure}[ht!]
\centerline{\psfig{file=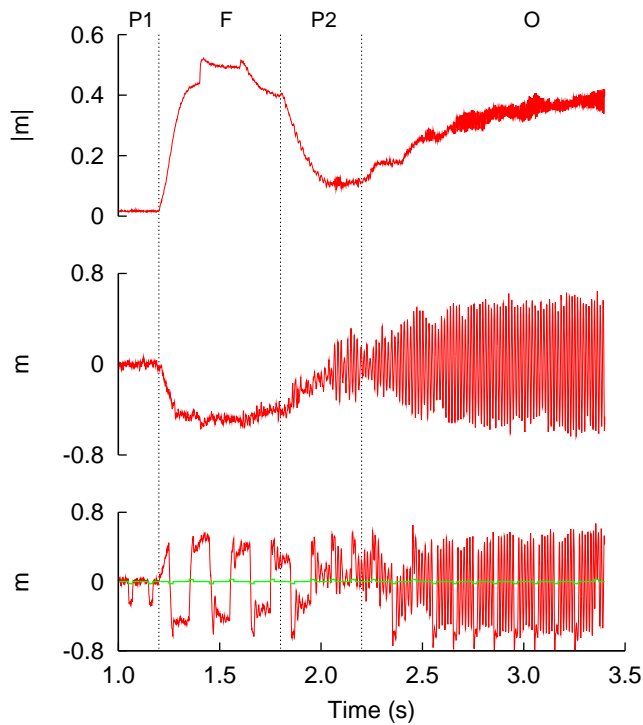, width=8.9cm}}
\caption{Time series for the overlap
function, $m,$ at $T=0.22$ (horizontal dotted line in figure \protect\ref%
{figure4}) as one increases the value of $\protect\tau _{\text{fac}}$ in
order to visit the different regimes (separated here by vertical lines). The
simulations started with $\protect\tau _{\text{fac}}=1$ at $t=1$ and $%
\protect\tau _{\text{fac}}$ was then increased by 10 units every 200 ms. The
bottom graph corresponds to a case in which the system is under the action
of an external stimulus (as described in the main text). The middle graph
depicts an individual run when the system is without any stimulus, and the
top graph corresponds to the average of $\left\vert m\right\vert $ over 100
independent runs of the unperturbed system.}
\label{figure6}
\end{figure}

\begin{figure}[ht!]
\centerline{\psfig{file=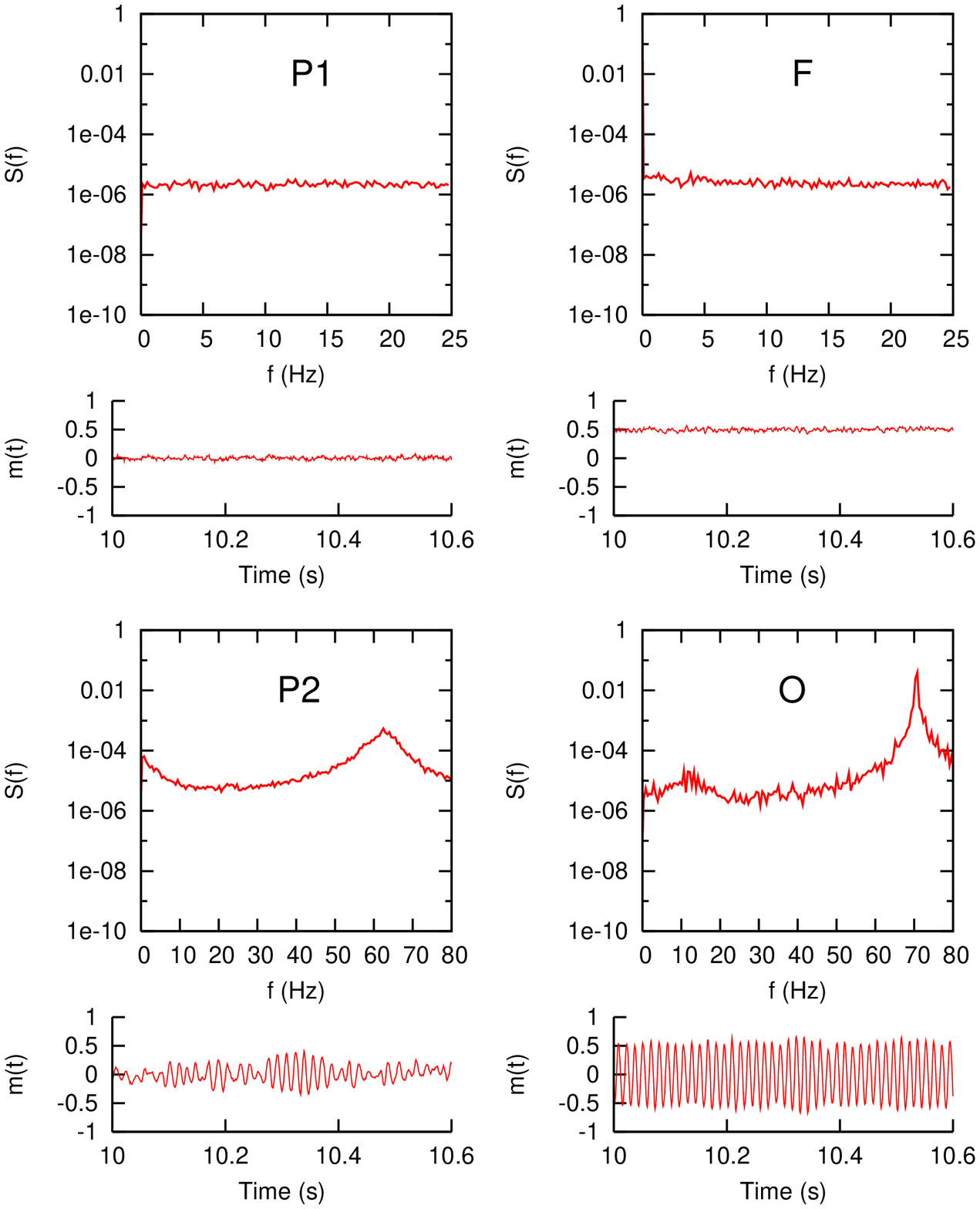,width=9.1cm}}
\caption{Spectral
analysis of the cases in figure \protect\ref{figure6}. On top of each of the
small panels, which show typical time series for $\protect\tau _{\text{fac}%
}=2,$ 20, 50 and 100, respectively, from top to bottom and from left to
right, the square panels show the corresponding power spectra. Details of
the simulations as in figure \protect\ref{figure6}.}
\label{figure7}
\end{figure}

In order to deepen further on some aspects of the system behavior, we
present in figures \ref{figure6} and \ref{figure7} a detailed study of
specific time series. The middle graph in figure \ref{figure6} corresponds
to a simulation of the system evolution for increasing values of $\tau _{%
\text{fac}}$ as one describes the horizontal line for $T=0.22$ in figure \ref%
{figure4}. The system thus visits consecutively the different regions
(separated by vertical lines) as time goes on. That is, the simulation
starts with the system in the stable \textit{paramagnetic} phase, denoted P1
in the figure, and then successively moves by varying $\tau _{\text{fac}}$
into the stable \textit{ferromagnetic} phase F, into another paramagnetic
phase, P2, and, finally, into the oscillatory phase O.

We interpret that the observed behavior in P2 is due to competition between
the facilitation mechanism, which tries to bring the system to the
fixed--point attractors, and the depression mechanism, which tends to
desestabilize the attractors. The result is a sort of intermittent behavior
in which oscillations and convergence to a fixed point alternates, in a way
which resembles (but is not) chaos. The top graph in figure \ref{figure6},
which corresponds to an average over independent runs, illustrates the
typical behaviour of the system in these simulations; the middle run depicts
an individual run.

Further interesting behavior is shown in the bottom graph of figure \ref%
{figure6}. This corresponds to an individual run in the presence of a very
small and irregular external stimulus which is represented by the (green)
line around $m=0.$ This consist of an irregular series of positive and
negative pulses of intensity $\pm 0.03\xi ^{1}$ and duration of 20 ms. In
addition to a great sensibility to weak inputs from the environment, this
reveals that increasing facilitation tends to significantly enhance the
system response.

Figure \ref{figure7} shows the power spectra of typical time series such as
the ones in figure \ref{figure6}, namely, describing the horizontal line for 
$T=0.22$ in figure \ref{figure4} to visit the different regimes. We plot
here time series $m\left( t\right) $ obtained, respectively, for $\tau _{%
\text{fac}}=2,$ 20, 50 and 100 and, on top of each of them, the
corresponding spectra. This reveals a flat, white--noise spectra for the P1
phase and also for the stable fixed--point solution in the F regime.
However, the case for the intermittent P2 phase depicts a small peak around
65 Hz. The peak is much sharper and it occurs at 70 Hz in the oscillatory
case.

\section{Conclusion}

We have shown that the dynamical properties of synapses have profound
consequences on the behaviour, and the possible functional role, of
recurrent neural networks.
Depending on the relative strength of the depression, the facilitation and the
noise in the network, one observes attractor dynamics to one of the
stored patterns, non-retrieval where the neurons fire largely at
random in a fashion that is uncorrelated to the stored memory
patterns, or switching where none of the stored patterns is stable and
the network switches rapidly between (the neighborhoods of) all of them. These three
behaviours were also observed in our previous work where we studied
the role of depression. 

The particular role of facilitation is the following.
The transitions between these possible phases are controlled by 
two facilitation parameters, namely, $\tau_{ \mathrm{fac}}$ and $U.$ 
Analysis of the oscillatory phase reveals
that the frequency of the oscillations, as well as the maximum
retrieval during oscillations increase when the degree of facilitation
increases. That is, facilitation favours in the model a faster access to the
stored information with a noticeably smaller error. This suggests that
synaptic facilitation might have an important role in short--term memory
processes.

There is
increasing evidence in the literature that similar jumping
processes could be at the origin of the animals ability to adapt and rapidly
response to the continuously changing stimuli in their environment. We
therefore believe that the network behaviour that is the consequence
of dynamic synapses as presented in this paper may have important
functional implications.

\section*{Acknowledgments}

This work was supported by the \textit{MEyC--FEDER} project FIS2005-00791, 
 the \textit{Junta de Andaluc\'{\i}a} project FQM 165 and the EPSRC-funded COLAMN project Ref. EP/CO 10841/1.
 We thank useful
discussion with Jorge F. Mej\'{\i}as.

\end{document}